\documentclass[pre,preprint,showpacs]{revtex4}
\usepackage{graphicx}

\begin{document}

\title{A random walker on a ratchet potential: Effect of a non
Gaussian noise}

\author{Sergio E. Mangioni\cite{conicet}}
\affiliation{Departamento de F\'{\i}sica, FCEyN,\\
Universidad Nacional de Mar del Plata\\
De\'an Funes 3350, 7600 Mar del Plata, Argentina}

\author{Horacio S. Wio}
\affiliation{Instituto de F\'{\i}sica de Cantabria, \\ Universidad
de Cantabria and CSIC,\\ Avda. Los Castros s/n, E-39005 Santander,
Spain }

\begin{abstract}
We analyze the effect of a colored non Gaussian noise on a model of
a random walker moving along a ratchet potential. Such a model was
motivated by the transport properties of motor proteins, like
kinesin and myosin. Previous studies have been realized assuming
white noises. However, for real situations, in general we could
expect that those noises be correlated and non Gaussian. Among other
aspects, in addition to a maximum in the current as the noise
intensity is varied, we have also found another optimal value of the
current when departing from Gaussian behavior. We show the relevant
effects that arise when departing from Gaussian behavior,
particularly related to current's enhancement, and discuss its
relevance for both biological and technological situations.
\end{abstract}


\maketitle

\section{Introduction}

Noise induced transport by Brownian motors or ``ratchets" has
attracted the attention of an increasing number of researchers due
to its biological interest as well as its potential technological
applications. Since the pioneering works, besides the built-in
ratchet-like bias and correlated fluctuations, several different
aspects have been studied, such as tilting \cite{tilt1,tilt2} and
pulsating \cite{puls} potentials, velocity inversions
\cite{tilt1,invers}, etc. There are relevant reviews
\cite{rev01,rev02} indicating the biological and/or technological
motivations for the study of ratchets as well as showing the state
of the art.

Among other aspects, ratchets has been used to explain the
unidirectional transport of molecular motors within a biological
realm \cite{rev01,rev02}. Among the different motor proteins,
kinesin has attracted considerable attention motivated by
experimental results in which the dynamical details of its motion
can be measured \cite{kin01,kin02,kin03}. Kinesin is a protein with
two heads that performs a walk along the microtubule inside cells.
Motivated by these experimental results, several researchers have
introduced diverse models in order to understand the particular form
of walking of kinesin \cite{kin04}. Usually those models consider a
walker moving along an asymmetric ratchet potential mediated by
noise. This walker has two feet that are represented by two
particles nonlinearly coupled through a bistable potential. Among
those models, there is a recent one introduced in \cite{mat01,mat02}
including all the above indicated ingredients, where the walker
moves along a track formed by an asymmetric potential, being
subjected to two independent white noise sources acting on each of
the two particles and to a common external harmonic force. It was
observed that the current $J$ as function of noise intensity $D_{w}$
presents a maximum, conforming another example of the constructive
role of noise.

Recent studies on the effect of a non Gaussian noise on several
noise induced phenomena, have shown the existence of strong effects
on the system`s response. In those studies, opposing to the most
usual cases that only consider Gaussian white noises, such an effect
was analyzed in stochastic resonance, noise induced transitions,
``standard" ratchets, etc \cite{qRE01,qRE02,qRE03,qRE04,qRE05}. This
form of noise was motivated by the nonextensive statistical
mechanics \cite{tsa01,tsa02}. In this work we want to analyze the
effect of this form of colored and non Gaussian noise over the
kinesin ratchet model introduced in Refs. \cite{mat01,mat02}. For
this purpose, we used a mean field approximation and exploited a
recently developed technique \cite{qRE05}. Through the variation of
a parameter $q$, this form of noise offers the possibility of
analyzing the departure from Gaussian behavior (corresponding to $q
=1$). Since subtle change of environment conditions can produce
drastic changes in biological process, such effects could be very
relevant. It is worth here indicating that a related mean field
approximation was introduced in \cite{nori1}.

The organization of the paper is as follows. In the next Section we
show the mean-field approximation and how reliable it is, based on
the good agreement with known results for the Gaussian case. The
third Section presents the statistical properties of the non
Gaussian noise, and the form of the current that results after
applying the \textit{effective Markovian approximation} in the
present problem. In the following Section we present the results
when we depart from the Gaussian behavior, that is varying the
parameter $q$. Finally, in the last Section we present some
conclusions.

\section{\label{model}The mean-field approximation and its accuracy:
Gaussian case}

The stochastic dimensionless differential equations for the two
particles in the overdamped regime, whose coordinates are indicated
by $x$ and $y$, are \cite{mat01}
\begin{eqnarray*}
\label{eqdifstxy}
    &&\dot{x}=-\partial_x V(x) -\partial_x V_{b}(x-y)+
    \sqrt{2D_{w}} \xi_{1}(t) + A \sin \Omega t,
    \\
    && \dot{y}=-\partial_y V(y) -\partial_y V_{b}(x-y)+
    \sqrt{2D_{w}} \xi_{2}(t) + A \sin \Omega t,
\end{eqnarray*}
where $V(x)$ is the dimensionless ratchet potential
\begin{equation}
\label{V} V(x)=C+U_{R}\,[ \sin(2\pi(x-x_{0}))- 0.25
\sin(4\pi(x-x_{0})) ]
\end{equation}
The constant  $C=U_{R}[\sin(2\pi x_{0})- 0.25 \sin(4\pi x_{0})]$ was
chosen in such a way that $V(0)=0$. The constant $x_{0}$ is
introduced in order to center the minima of the periodic potential
on integer values. The dimensionless amplitude of the ratchet
potential is indicated by $U_{R}$.

The dimensionless bistable potential $V_{b}(x-y)$, that represents
the nonlinear coupling between the two particles, is given by
\begin{equation}
\label{Vb} V_{b}(x-y)=U_{b}\, \left[1+
\frac{(x-y)^{4}}{l^{4}}-2\frac{(x-y)^{2}}{l^{2}} \right],
\end{equation}
where $U_{b}$ is the dimensionless amplitude of this bistable
potential and $2l$ is the distance between the two minima.

In the original model, it is assumed that $\xi_{1}(t)$ and
$\xi_{2}(t)$ are Gaussian white noises with zero mean and
correlation $\langle \xi_{i}(t) \xi_{i}(t') \rangle= 2D_{w}
\delta_{i,j} \delta(t-t')$, with $D_{w}$ the intensity of the
statistically independent noises. In our model, as indicated at the
introduction, we assume they are non Gaussian colored noises, with
characteristics that we will briefly indicated later. Clearly, the
model also considers an external harmonic force.

In \cite{mat02} this model was analytically solved. However, in
order to consider the non Gaussian colored noise case, we need to
introduce a different approach. For that purpose, we consider a mean
field approximation (MFA) \cite{mfa01}. For the present case, such a
MFA consist in the following approximation for $V_{b}(x- y)$
\begin{equation}
\label{Vb2} V_{b}(x- M)=U_{b}\, \left[1+
\frac{(x-M)^{4}}{l^{4}}-2\frac{(x-M)^{2}}{l^{2}} \right],
\end{equation}
with $M = <y>$, and clearly we have $<x> = <y>$. After applying the
MFA, the equations for both variables $x$ and $y$ results to be of
the same form. Hence, we can reduce the problem to a single equation
describing the model system
\begin{equation}
\label{EQDIFST} \dot{x}=-\partial_x V_{eff}(x,M,t) + \sqrt{2D_{w}}
\xi(t),
\end{equation}
where $V_{eff}(x,M,t)=V(x) + V_{b}(x-M) - x A \sin \Omega t$.

In order to test the MFA, we start analyzing the Gaussian white
case, and show that it gives similar qualitative results than the
numerical simulations and the analytical solution for the original
model \cite{mat01,mat02}. Hence, in Eq. (\ref{EQDIFST}), we start
assuming that $\xi(t)$ is a Gaussian white noise with the same
behavior than the $\xi_i(t)$'s.

Considering an adiabatical approximation in order to decouple the
external harmonic force from the rest, the corresponding
Fokker-Planck equation can be solved assuming periodic boundary
conditions. Even though $V_{eff}(x,M,t)$ is non periodic [due to
$V_{b}(x - M)$ being non periodic], we can still assume such a
periodic behavior since both feet are never too far away from each
other, the strong slope of the attractive potential preventing it to
occur. In Fig. 1 it is shown $V_{b}(x)$ vs. $x$ with and without the
indicated approximation. In Fig. 2 we show $V_{eff}(x)$ vs. $x$,
again with and without the approximation. It is apparent that the
curves looks quite alike.

\begin{figure}
\centering \resizebox{.8\columnwidth}{!}{\includegraphics{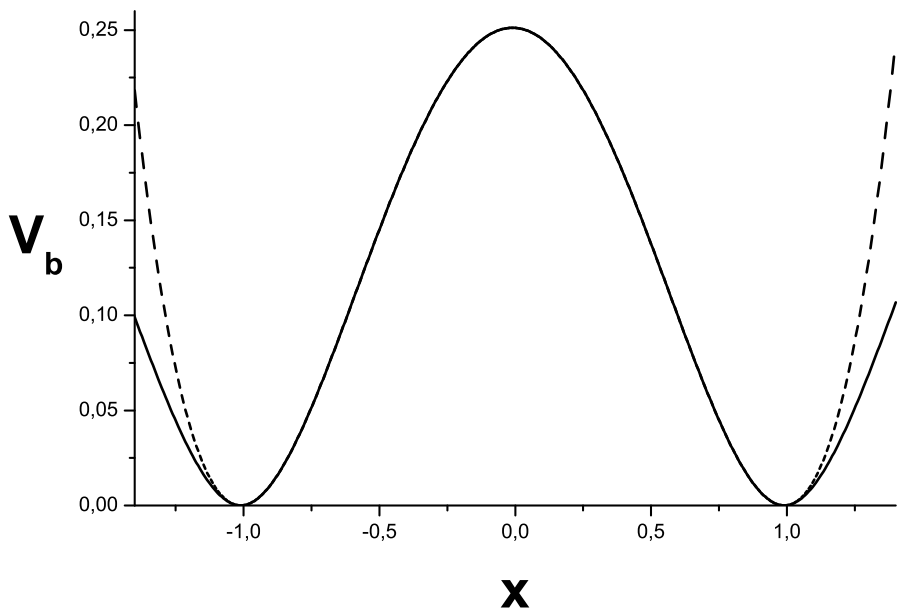}}
\caption{The bistable potential $V_{b}$ vs. $x$. The continuous line
corresponds to the exact form of the potential, while the dashed
line is for the approximate one. The parameters are $U_{b}=0.2512$
and $l=1$.} \label{Fig1}
\end{figure}

\begin{figure}
\centering \resizebox{.8\columnwidth}{!}{\includegraphics{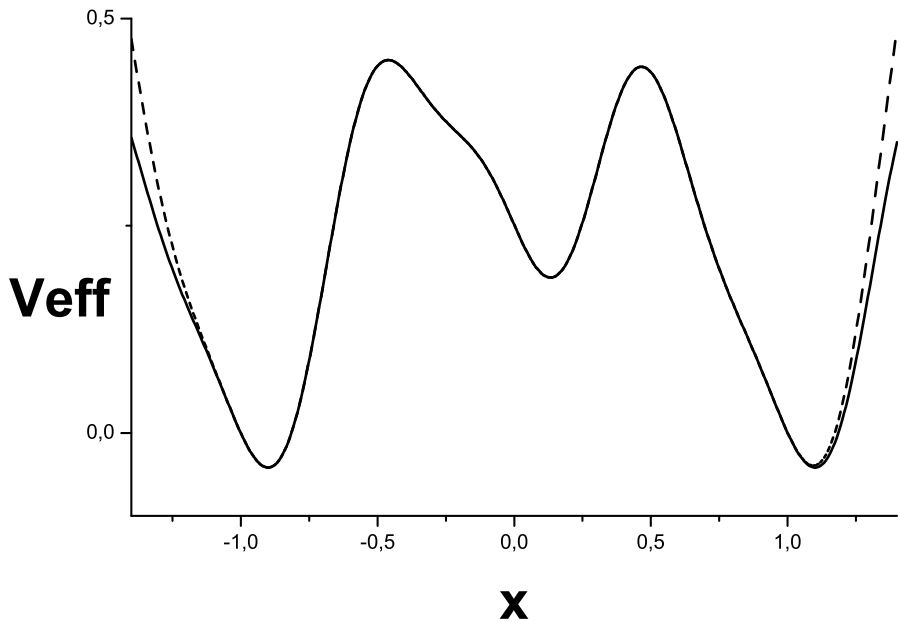}}
\caption{Effective potential $V_{eff}$  vs. $x$. The continuous line
corresponds to the exact form of the potential, while the dashed
line is for the approximate one. The parameters are $A=1$, $l=1$,
$U_{r}=0.16$, $U_{b}=0.2512$ and $t=0$.} \label{Fig2}
\end{figure}

Considering the adiabatical approximation, the corresponding
stationary solution of the Fokker-Planck equation is \cite{gard}
\begin{equation}
\label{eqdifstxy} P^{st}(x,M)=\frac{H(x)}{N
\,\sqrt{2D_{w}}}\,e^{-V_{eff}(x,M)/D_{w}},
\end{equation}
with
\begin{equation}
H(x)=\frac{1}{\sqrt{2D_{w}}}\,\int^{x+L}_{x} du \,
e^{V_{eff}(u,M)/D_{w}},
\end{equation}
$N$ a normalization constant, and $L=2\pi$ the period. Hence,
numerically solving
\begin{equation}
\label{eqcm} \int^{L/2}_{-L/2} dx \, x \, P^{st}(x,M) = M = <x>
\end{equation}
we can obtain the current $J(t)$ \cite{gard} as
\begin{equation}
\label{Jt} J(t)=\frac{1}{2N} \left[1 -
e^{(V_{eff}(L)-V_{eff}(0))/D_{w}}\right],
\end{equation}
and the net current is obtained as
\begin{equation}
\label{J} J=\frac{1}{T} \int^{T}_{0} J(t)\, dt,
\end{equation}
with $T= \frac{2\pi}{\Omega}$.

In order to check these results in Fig. 3 we depict $J$ vs. $D_{w}$.
The curve presents a maximum for an ``optimal" value of the noise
intensity, which is qualitatively similar to the one obtained in
\cite{mat01,mat02}. With the support offered by this agreement, we
are now in position to use this same approach for the non Gaussian
colored noise case.

\begin{figure}
\centering \resizebox{1.\columnwidth}{!}{\includegraphics{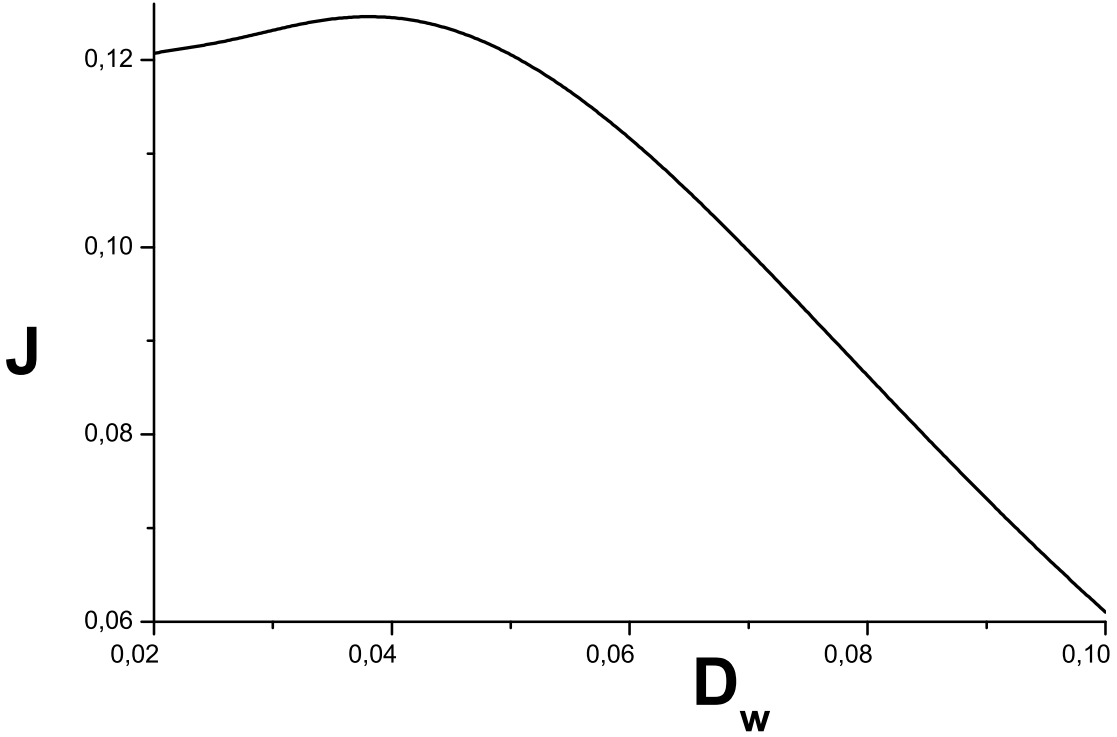}}
\caption{Net current $J$ for a gaussian white noise, as function of
$D_{w}$. The values of parameters are $A=1$, $l=1$, $U_{r}=0.16$ and
$U_{b}=0.2512$. } \label{Fig3}
\end{figure}

\section{\label{noise}Non Gaussian noise case}

\subsection{Statistical properties of the non Gaussian noise}

We consider now the non Gaussian colored noise case as in
\cite{qRE01}. Hence, in Eq. (\ref{EQDIFST})  $\xi(t)$ is a noise
with a dynamics described by the following Langevin equation
\begin{equation}
\label{se}
 \dot{\xi(t)}=-\frac{1}{\tau} \frac{d}{d\xi)}V_{q}(\xi) +
\frac{1}{\tau} \eta(t),
\end{equation}
with $\eta(t)$ a white Gaussian noise, and
\begin{equation}
\label{Vq} V_{q}(\xi) = \frac{D_{w}}{\tau(q-1)}\ln \left[
1+\frac{\tau}{D_{w}(q-1)\frac{\xi^{2}}{2}} \right].
\end{equation}
For $q=1$, the process $\xi$ coincides with the Ornstein-Uhlenbeck
(OU) one (with a correlation time equal to $\tau $), while for
$q\neq1$ it departs from the Gaussian behavior. For $q<1$ the
stationary probability distribution (spd) has a bounded support,
with a cut-off given by
$\|\xi\|=\omega\equiv[(1-q)\tau/(2D_{w})]^{-\frac{1}{2}}$, with a
form given by
\begin{equation}
\label{Pq1} P_{q}(\xi) = \frac{1}{Z_{q}}\,\left[
1-(\frac{\xi}{\omega})^{2} \right]^{\frac{1}{1-q}},
\end{equation}
for $\|\xi\|<\omega$ and zero for $\|\xi\|>\omega$ ($Z_{q}$ is a
normalization constant). Within the range $1<q<3$, the spd is given
by
\begin{equation}
\label{Pq2} P_{q}(\xi) = \frac{1}{Z_{q}}\, \left[
1+\frac{\tau(q-1)\xi^{2}}{2D_{w}} \right]^{\frac{1}{1-q}},
\end{equation}
for $-\infty<\xi<\infty$, and decays as a power law (that is, slower
than a Gaussian distribution) for $\xi \to \infty$. Finally, for
$q>3$, this distribution can not be normalized. Note that keeping
$D_{w}$ constant, when increasing $q$ the dispersion of the
distribution also increases. In \cite{qRE01} the second moment of
the distribution was obtained. This moment is related to the
intensity of the non Gaussian noise, and is given by
\begin{equation}
\label{D} D_{ng}=<\xi^{2}>=\frac{2D_{w}}{\tau(5-3q)},
\end{equation}
that diverges for $q \geq 5/3$. For $\tau_{ng}$, the correlation
time of the process $\xi(t)$, that was defined in detail in
\cite{qRE01}, it was not possible to find an analytical expression.
However, it is known \cite{qRE01} that for $q \to 5/3$ it diverges
as $(5-3q)^{-1}$. In \cite{qRE01} $\tau_{ng}$ vs. $q$ was
numerically calculated for the range $0.5 < q < 5/3$. The following
analytical approximation, accurate within the indicated range, was
found
\begin{equation}
\label{Dgn}
 \tau_{ng}=\frac{2\,\tau}{(5-3q)}\,[1+4(q-1)^{2}].
\end{equation}

\subsection{\label{JT}Current in the non Gaussian case}

The Fokker-Planck equation for the present case, that is the system
described by the Langevin equation indicated by Eq. (\ref{EQDIFST})
with $\xi(t)$ the non Gaussian noise, has a known structure. After
applying the \textit{effective Markovian approximation} as exploited
in \cite{qRE01,qRE03,qRE05}, such an equation has the form
\begin{equation}
\label{FPE} \partial_{t} P(x,t)=\partial_{x} [A(x)P(x,t)] +
\frac{1}{2} \partial^{2}_{x} [B(x)P(x,t)],
\end{equation}
where
\begin{equation}
A(x)=\frac{V_{eff}'}{A_1(x)+A_2(x)}
\end{equation}
with
\begin{eqnarray*}
    &&A_1(x)=\frac{1-(\tau/2D_{w})(q-1)V_{eff}'^{2}}
    {1+(\tau/2D_{w})(q-1)V_{eff}'^{2}}
    \\
    &&A_2(x)=\tau V_{eff}''[1+(\tau/2D_{w})(q-1)V_{eff}'^{2}];
\end{eqnarray*}
and
\begin{equation}
B(x)=D_{w}\, \left[ \frac{[1+(\tau/2D_{w})(q-1)V_{eff}'^{2}]^{2}}
    {B_1(x)+B_2(x)} \right]^{2}
\end{equation}
with
\begin{eqnarray*}
    &&B_1(x)=\tau V_{eff}''\, \left[1+(\tau/2D_{w})(q-1)V_{eff}'^{2} \right]^{2}
    \\
    &&B_2(x)=\left[1-(\tau/2D_{w})(q-1)V_{eff}'^{2} \right].
\end{eqnarray*}
The prima indicates a derivative respect to $x$.

The corresponding stationary solution of the indicated Fokker-Planck
equation is \cite{gard}
\begin{equation}
\label{s2} P^{st}(x)=\frac{e^{-\phi(x)}}{N S(x)}
    \int^{x+L}_{x} du \frac{e^{\phi(u)}}{S(u)},
\end{equation}
with $N$ a normalization constant, and
\begin{equation}
\label{fi2} \phi(x)=-\int^{x}_{0}du
\frac{R(u)}{S^{2}(u)},
\end{equation}
where $S(x)=\sqrt{B}$ and $R(x)=-A-\frac{1}{2}S\,S'$. Then, using
Eqs. (\ref{eqcm}), as (\ref{Jt}) and (\ref{J}) we can obtain the
current $J$.

\section{\label{JQ}Results}

In what follows we present several results for the case of
submitting the system to a non Gaussian noise, mainly restricting
ourselves to the range $0.5 < q < 1.5$, which is the range where our
evaluation is, in principle, valid. In all cases, except we
indicated something different, we adopted $\tau_{ng}=0.01 \pi$,
$A=1$, $l=1$, $\Omega=0.5$, $U_{r}=0.16$ and $U_{b}=0.2512$.

In Fig. 4 we show the dependence of $J$ on $q$, for different values
of $D_{ng}$. Looking at the left part of the picture, we identify
the curves, from top to bottom, with the values $D_{ng}=0.02$,
$0.038$, $0.06$, $0.07$, $0.085$, $0.1$ and $0.13$. The entanglement
between the dependence on $q$ and $D_{ng}$ is apparent. For low
values of $D_{ng}$ ($D_{ng}\sim 0.02$) the behavior of $J$
corresponds to an initial plateau for $q<1$ and a fast (step like)
decrease for $q>1$. When $D_{ng}$ increases ($D_{ng}\gtrsim 0.06$),
$J(q)$ adopts a bell's shape, with a maximum that shifts to larger
values of $q$ for increasing $D_{ng}$. It is worth noting that for
large values of $D_{ng}$, the relevant values for the current will
only occur when $q$ takes values within the indicated bell-like
region.

\begin{figure}
\centering \resizebox{.8\columnwidth}{!}{\includegraphics{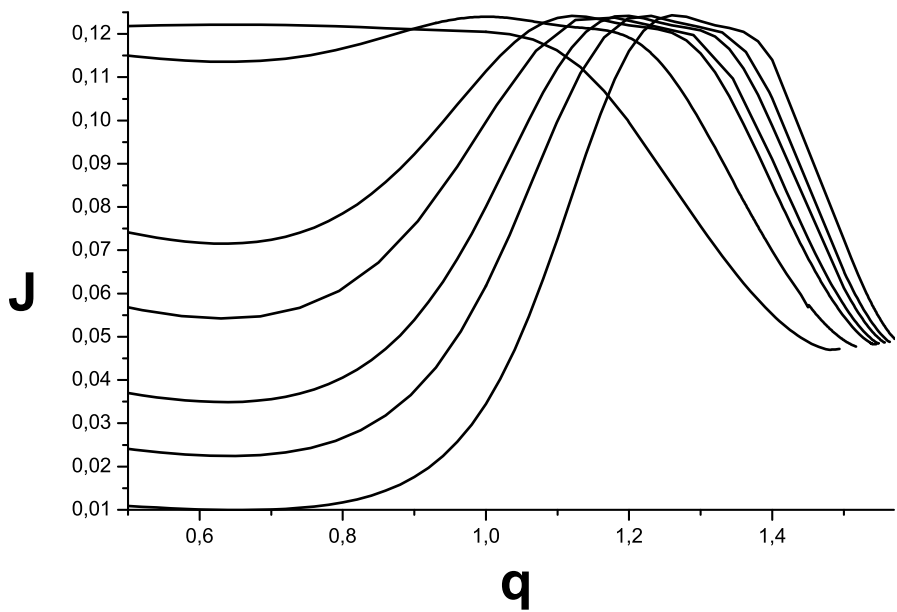}}
\caption{Net current $J$ vs. $q$, for the range $.5<q<1.5$. Looking
at the left part of picture, we identify from top to bottom $D_{ng}
= 0.02$, $0.038$, $0.06$, $0.07$, $0.085$, $0.1$ and $0.13$. Other
parameters are: $\tau_{ng}=0.01\pi$, $A=1$, $l=1$, $U_{r}=0.16$ and
$U_{b}=0.2512$.} \label{Fig4}
\end{figure}

As in previous studies on the effect of non Gaussian noise on other
systems \cite{qRE01,qRE02,qRE03,qRE04,qRE05} we conclude that the
observed change of behavior is associated with the process $\xi(t)$,
that is with the departure from Gaussian behavior ($q=1$) and the
associated changes in $P_{q}(\xi)$. As indicated before, for $q<1$
there is a cut-off that prevents that $\xi(t)$ reaches large values,
while for $q>1$ $\xi(t)$ can reach very large values. For small
values of $D_{ng}$ it is expected that the current $J(q)$ will not
be very sensitive to such a cut-off, as the probability for $\xi(t)$
to reach those cut-off values is very small. In opposition, for
large values of $D_{ng}$ we expect that $J(q)$ becomes highly
sensitive to the cut-off. It seems to be the case as $J$ varies very
slowly with $q$ (for $q<1$) when $D_{ng}$ is small, while for large
values of $D_{ng}$ just the opposite occurs.

However, for $q>1$, it is the tail of the spd  that mainly
contributes to build the current \cite{qRE05}. As indicated above,
for $q<1$ (that is, for the case without tail), and large values of
$D_{ng}$, $J$ drastically decreases when $q$ decreases. Let us
select one of the curves for $J$ with a bell shape, and look into
the spd for values of $q$ below, at, and above of the maximum of
that curve. Such analysis is shown in Fig. 5. We have chosen
$D_{ng}=0.085$, $q=0.94$ (below), $q=1.2$ (at the maximum) and
$q=1.4$ (above). We observe that for the case of the maximum, the
spd is symmetric with two peaks at both ends of the space period.
However, when departing from such ``optimal" condition, the spd
looses its symmetry. In one hand, when $q$ is at the left of the
maximum ($q=0.94$) a new peak emerges at the central zone of the
period. On the other hand, when $q$ is to right of the maximum,
there are two peaks that arise between the previously indicated
ones. In both cases, the prize of these additional peaks is that the
spd decreases at both ends, indicating a reduction of $J$. The same
behavior was also observed in all cases where $J(q)$ adopts a bell's
shape. We also compare the spd behavior for the case when $J(q)$ has
a step-like shape, and note that for values of $q$ within the
plateau, the spd is like the one corresponding to the case of the
maximum of $J(q)$ (Fig. 3, for $q=1.2$), while for values of $q$
beyond the step, the spd is analogous to the one corresponding to a
value of $q$ located at the right of the maximum ($q=1.4$). It seems
that the current reaches a maximum when one feet follows the other,
and their separation is kept constant, and equal to a period. We
have also analyzed the Gaussian case ($q=1$) for different values of
$D_{ng}$, with results that essentially reproduces those of Fig. 5.
In short, for small $D_{ng}$ the curves are analogous to those
obtained for large $q$ (having four peaks), while for large $D_{ng}$
the shapes resembles the ones for small $q$ (having three peaks).

\begin{figure}
\centering \resizebox{.8\columnwidth}{!}{\includegraphics{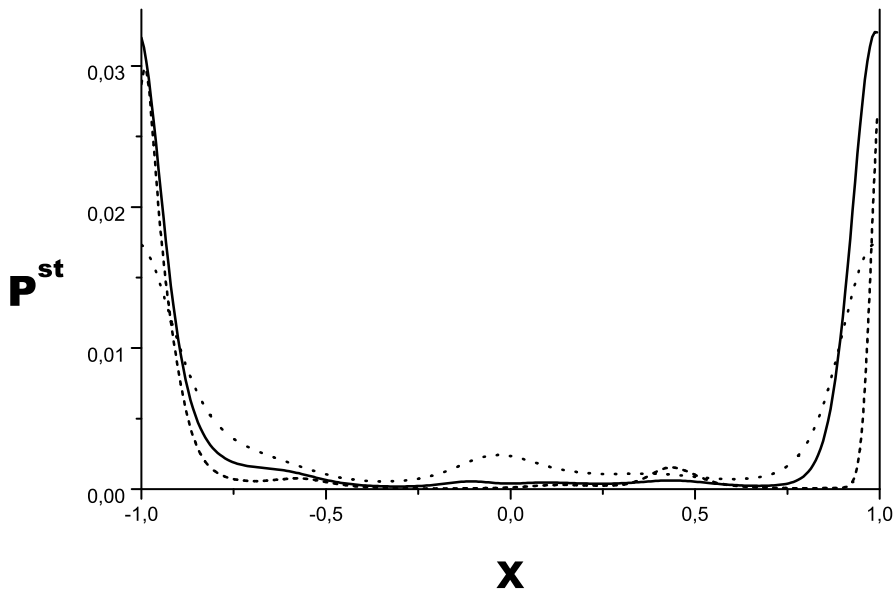}}
\caption{Stationary probability distribution $P^{st}$ for $q$ below,
at, and after of maximum $J$'s value, with $D_{ng}=0.085$ and: dot
line for $q=0.94$, solid line for $q=1.2$ and dash line for $q=1.4$.
Other parameters are: $\tau_{ng}=0.01\pi$, $A=1$, $l=1$, $U_{r} =
0.16$ and $U_{b}=0.2512$.} \label{Fig5}
\end{figure}

Figure 6 depicts $J(t)$ --the current before its time averaging-- as
function of time, for a fixed $D_{ng}$ and different values of $q$.
It is apparent that the current adopts both, positive and negative
values along the period of the external forcing. We observe that for
$q<q_{max}$ (where $J(q_{max})$ reaches its largest values) there is
a balance between positive and negative current's values yielding,
in average, a small net current. When $q$ increases, the negative
values of the current reduces, the indicated balance is also
reduced, and the net current increases. For even larger values of
$q$ ($q>q_{max}$), the positive value of the current is reduced,
while the negative one tend to zero, and the net current decreases.

\begin{figure}
\centering \resizebox{.8\columnwidth}{!}{\includegraphics{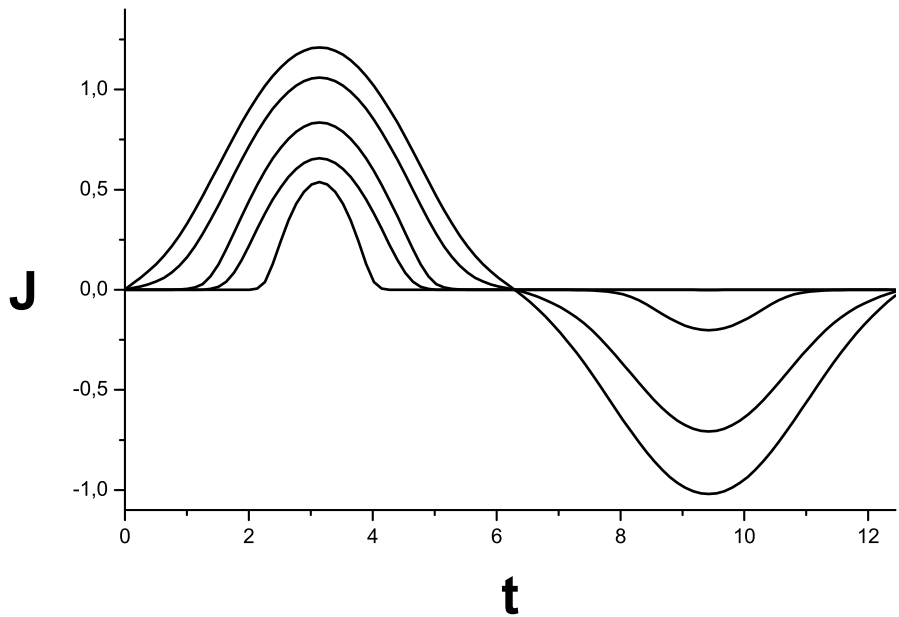}}
\caption{$J(t)$, the current before the time averaging, vs. $t$, for
$D_{ng}=0.085$ and different values of $q$'s. Adopting as reference
the left part of the figure, from top to bottom we have $q=0.3$,
$1.$, $1.2$, $1.4$ and $1.55$. Other parameters are:
$\tau_{ng}=0.01\pi$, $A=1$, $l=1$, $U_{r}=0.16$ and $U_{b}=0.2512$.}
\label{Fig6}
\end{figure}

We finally analyze the system's behavior with $\tau_{ng}$. Firstly,
we observed that inside the studied range, $J$ varies almost
linearly with $\tau_{ng}$. Since the corresponding slope change its
sign with $q$ and $D_{ng}$, we choose to look into the behavior of
$dJ/d\tau_{ng}$. Hence, in Fig. 7 we show $dJ/d\tau_{ng}$ vs. $q$
for different values of $D_{ng}$, while in Fig. 8 we show
$dJ/d\tau_{ng}$ vs. $D_{ng}$ for different values of $q$. The
sensitivity of $dJ/d\tau_{ng}$ against variations of $q$ and
$D_{ng}$ is apparent, and the changes in the slope's sign are
clearly seen.

\begin{figure}
\centering \resizebox{.8\columnwidth}{!}{\includegraphics{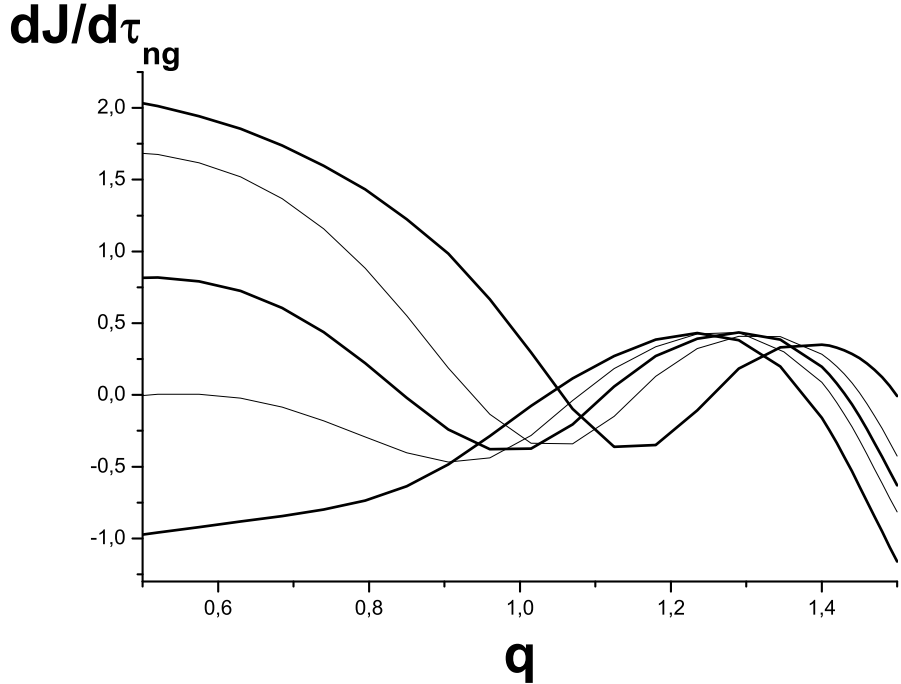}}
\caption{$dJ/d\tau_{ng}$ vs. $q$ for different $D_{ng}$. Considering
the left part of the figure, we identify from top to bottom,
$D_{ng}=0.1$, $0.05$, $0.038$, $0.03$ and $0.02$. Other parameters
are: $\tau_{ng}=0.01\pi$, $A=1$, $l=1$, $U_{r}=0.16$ and
$U_{b}=0.2512$.} \label{Fig7}
\end{figure}

\begin{figure}
\centering \resizebox{.8\columnwidth}{!}{\includegraphics{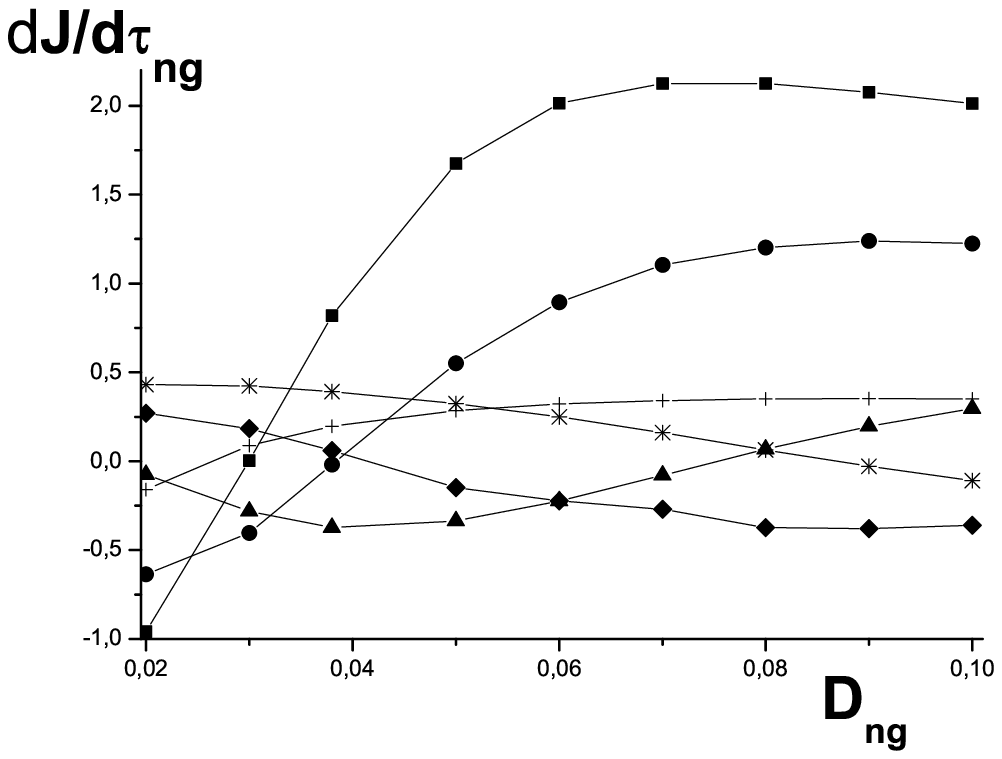}}
\caption{$dJ/d\tau_{ng}$ vs. $D_{ng}$ for different values of $q$:
squares $q$=0.52, circles $q$=0.85, triangles $q$=1.015, diamonds
$q$=1.125, stars $q$=1.235 and  crosses $q$=1.4. Other parameters
are: $\tau_{ng}=0.01\pi$, $A=1$, $l=1$, $U_{r}=0.16$ and
$U_{b}=0.2512$.} \label{Fig8}
\end{figure}

\section{\label{C}Conclusions}

In line with some recent work \cite{qRE01,qRE02,qRE03,qRE04,qRE05},
we have here analyzed the effect of a colored and non Gaussian noise
over the kinesin ratchet model introduced in \cite{mat01,mat02}. As
was discussed in \cite{qRE01}, there are strong experimental
evidences that noise sources within biological systems can not be in
general Gaussian \cite{biol}. This fact gives strong support to the
study of the effect of non Gaussian noises within a biological
motivated context.

In order to make an almost analytical treatment, we have here
exploited a mean-field like approach, that shows a nice agreement
when it is tested  against known results \cite{mat01,mat02} for the
Gaussian case ($q=1$). Hence, we can trust the results we have
obtained when varying the parameter $q$. For the adopted form of
noise, such variation offers the possibility of analyzing the
departure ($q \neq 1$) from the Gaussian behavior ($q =1$).

When analyzing the dependence of the current $J$ on $D_{ng}$ and
$q$, we have confirmed the existence, for a fixed value of $q$, of a
maximum or ``optimal" current as function of $D_{ng}$. What is new,
is that for fixed $D_{ng}$, there is also an optimal value of $q$
yielding a maximum value of the current. More, in general such a
value corresponds to a non Gaussian situation ($q \neq 1$). Such a
results could be understood from the behavior of the spd for
different values of $q$, as well as from the current's time
dependence (that is, before the time averaging). Regarding the
dependence of $J$ on $\tau _{ng}$, we have seen that it is linear.
However, analyzing its slope ($dJ/d\tau _{ng}$) it becomes apparent
the possibilities of even changing the sign of such a slope varying
both, $q$ for fixed $D_{ng}$, or $D_{ng}$ for fixed $q$.

The indicated results show us the richness of behavior that we can
found when departing from the Gaussian situation. It is apparent
that those results could be of relevance not only for biological
studies, but for technological applications as well. This point was
discussed in general in \cite{qRE04}, and particularly for the case
of ratchets in \cite{qRE05}. From a technological point of view,
such a noise source offers a variety of forms of controlling and/or
optimizing the transport process.

Clearly, the exploitation of an approach like the one used in
\cite{mat02}, based in a coordinate separation, could offer a more
transparent and better description of this system's physics. In
addition a numerical validation of the whole approach is required.
Also, another aspect that requires further analysis is the case when
$2l$ is non-commensurate with the unit cell of the ratchet
potential. All these aspects will be the subject of forthcoming work
\cite{manwio2}.

\begin{acknowledgments}
SEM acknowledges financial support from CONICET, Argentina, PIP
5072, and from UNMdP, EXA 338/06. HSW acknowledges financial support
from MEC, Spain, through Grant No. CGL2004-02652/CLI, and thanks the
European Commission for the award of a \textit{Marie Curie Chair}
during part of the development of this work.
\end{acknowledgments}

\end{document}